\documentclass[11pt]{article}

\usepackage{amsfonts}
\usepackage{amsbsy} 
\usepackage{epsfig}
\usepackage{latexsym}
\input amssym.def
\input amssym.tex
\def\a{{\mathcal{M}}}
\advance \topmargin by -\headheight
\advance \topmargin by -\headsep
\evensidemargin \oddsidemargin
\advance\hoffset by -3mm  
\advance\voffset by  8mm  
\def\bbbc{{\mathchoice {\setbox0=\hbox{$\displaystyle\rm C$}\hbox{\hbox
to0pt{\kern0.4\wd0\vrule height0.9\ht0\hss}\box0}}
{\setbox0=\hbox{$\textstyle\rm C$}\hbox{\hbox
to0pt{\kern0.4\wd0\vrule height0.9\ht0\hss}\box0}}
{\setbox0=\hbox{$\scriptstyle\rm C$}\hbox{\hbox
to0pt{\kern0.4\wd0\vrule height0.9\ht0\hss}\box0}}
{\setbox0=\hbox{$\scriptscriptstyle\rm C$}\hbox{\hbox
to0pt{\kern0.4\wd0\vrule height0.9\ht0\hss}\box0}}}}

\makeatletter
\@addtoreset{equation}{section}
\makeatother

\begin{document}
\hfuzz=100pt
\title{{\Large \bf{$\bbbc P^2$ and $\bbbc P^{1}$ Sigma Models in Supergravity:
Bianchi type IX Instantons and Cosmologies}}}
\author{\\M M Akbar\footnote{E-mail:M.M.Akbar@damtp.cam.ac.uk}\ \,\& P D D'Eath\footnote{E-mail:P.D.DEath@damtp.cam.ac.uk}  
\\
\\
Department of Applied Mathematics and Theoretical Physics,
\\ Centre for Mathematical Sciences,
\\ University of Cambridge, 
\\ Wilberforce Road,
\\ Cambridge CB3 0WA,
 \\ U.K.}

\maketitle
\begin{center}
DAMTP-2002-37
\end{center}
\begin{abstract}

We find instanton/cosmological solutions with biaxial Bianchi-IX
symmetry, involving non-trivial spatial dependence of the $\bbbc
P^{1}$- and $\bbbc P^{2}$-sigma-models coupled to gravity. 
Such manifolds arise in $N=1$, $d=4$ supergravity with supermatter
actions and hence the solutions can be embedded in supergravity. There
is a natural way in which the standard coordinates of these manifolds
can be mapped into the four-dimensional physical space. Due to its
special symmetry, we start with $\bbbc P^{2}$ with its corresponding
scalar Ansatz; this further requires the spacetime to be $SU(2) \times
U(1)$-invariant. The problem then reduces to a set of ordinary
differential equations whose analytical properties and solutions are
discussed. Among the solutions there is a surprising, special-family
of exact solutions which owe their existence to the non-trivial
topology of $\bbbc P^{2}$ and are in 1-1 correspondence with
matter-free Bianchi-IX metrics. These solutions can also be found by
coupling $\bbbc P^{1}$ to gravity. The regularity of these Euclidean
solutions is discussed -- the only possibility is bolt-type
regularity. The Lorentzian solutions with similar scalar Ansatz are
all obtainable from the Euclidean solutions by Wick rotation.
\end{abstract}

\section{Introduction}
\noindent
Because of their non-linearities exact solutions of the Einstein
equations, in vacuo or in the presence of a possible cosmological
constant, are usually obtained assuming a large degree of symmetry and
other simplifying features.  In the presence of matter fields,
especially non-linear ones, exact solutions are much rarer and are
known only in the simplest matter configurations. These solutions
however are the starting points for the full dynamical study of the
gravity-matter system.

Largely because of their possible role in inflationary cosmology,
cosmological solutions with scalars coupled to gravity have garnered
particular interest in both Lorentzian and Riemannian signatures. In
supergravity theories scalar fields arise naturally as sigma models
and hence the resulting system is more non-linear and difficult to
solve. In most work on the subject, however, authors tend to ignore
the geometry of the `target' manifold on which the scalar fields live,
with the consequence that it does not play a role in determining the
geometry of the spacetime and makes the system simpler; attention has
usually focussed much more on the form of the scalar potential. In this
paper, we will see that information of a different kind emerges if we
let the target manifold play a more active role.

For definiteness we consider the $\bbbc P^{1}$- and $\bbbc
P^{2}$-sigma-models coupled to gravity with a cosmological
constant. Such models can be embedded, for example, in $N=1$, $d=4$
supergravity with supermatter Lagrangian (see below).  In fact, they
arise naturally for the gauge group of $SU(3)$ and $SU(2)$ in $N=1$,
$d=4$ supergravity coupled to gauged supermatter Lagrangian. For
reasons of symmetry involving the cosmological models, we first
consider $\bbbc P^{2}$. With a non-trivial, but natural, Ansatz for
the scalars, the symmetry of spacetime naturally emerges to be that of
biaxial Bianchi-IX. We derive the general set of field equations and
obtain, among others, a set of special-case exact solutions which are
``deformations'' of the Bianchi-IX biaxial metrics satisfying the
Einstein-$\Lambda$ equations and are in 1-1 correspondence with
them. These solutions exist for both the Lorentzian and Euclidean
signatures. The Lorentzian solutions can all be obtained by Wick
rotation and hence we will be describing the Euclidean solutions and
their regularity in greater length. Some of the Euclidean solutions
can be extended over complete Riemannian manifolds and hence are
instanton solutions of the coupled system. We then show that this
special set of solutions can be obtained from the $\bbbc P^{1}$ sigma
models. All solutions that we will be reporting in this paper make
crucial use of the geometry of the target manifold.

This paper is arranged in the following way. In section 2 we discuss
the theoretical framework and the action; in section 3 we describe the
Ansatz for the scalar fields for the $\bbbc P^{2}$-sigma model and the
spacetime and deduce the field equations. In section 4 we describe
the solutions. In section 5 we discuss the regularity of the solutions
and show how these solutions can in fact be obtained for the $\bbbc
P^{1}$-sigma model as well. Finally, we conclude with comments on the
corresponding Lorentzian solutions.
\section{Actions}
\noindent
In this paper we will essentially find solutions corresponding to the
following (two) Euclidean actions:
\begin{equation}
I_{E} = -\int d^4 x \sqrt{g} \left(\frac{1}{2} R - 
g_{ij^{*}}{\partial}_{\mu}a^{i}{\partial}^{\mu}a^{*j} - \Lambda\right)\label{1}
\end{equation}
where $g_{\mu \nu}$ is the spacetime metric, with $g=\hbox{det}(g_{\mu
\nu})$, and $g_{ij^{*}}$ is the metric on the target (complex)
manifold, here the complex projective spaces $\bbbc P^2$ and $\bbbc
P^{1}$ respectively, and $\Lambda$ is an arbitrary cosmological
constant. The motivation for studying such sigma models comes from
supergravity as scalar fields in supergravity theories take their
values on K\"{a}hler manifolds.
\subsection{$N=1$, $d=4$ supergravity with supermatter action}
It is easy to check that the action (\ref{1}) is a valid truncated of
the full $N=1$, $d=4$ supergravity action whose bosonic part is
\cite{WB}
\begin{equation}
\!{\mathcal I}\!=\!-\!\!\int\!d^4 \!x\sqrt{g}\left(\!\frac{1}{2}R\!-\!g_{ij^{*}}{\partial}_{\mu}a^{i}{\partial}^{\mu}a^{*j}\!-\!\frac{1}{2}D_{a}D^{a}\!\!-\!\frac{1}{4}\mathrm{Re}(h_{ab})F^{(a)}_{\rho\sigma}F^{(a)\rho\sigma}\!+\!\!\frac{1}{4}\mathrm{Im}(h_{ab})F^{(a)}_{\rho\sigma}\!
\,^{*}\!\!F^{(a)\rho\sigma}\!\!-\!\!V_{F}(a^{i}\!\!, a^{\!*j})\!\right)\!\!.\label{2}
\end{equation}
Here $g_{ij^{*}}=\partial^{2}K/\partial a^{i}\partial a^{j^{*}}$ is
the K\"{a}hler metric which is derived from a K\"{a}hler potential
$K(a^{i},a^{*j})$ and
$F^{(a)}_{\mu\nu}=\partial_{\mu}v_{\nu}^{(a)}-\partial_{\nu}v_{\mu}^{(a)}$
are the Maxwell field strengths (with $U(1)$ gauge group for each
index $(a)$) and
\begin{equation}
V_{F}=e^{K}\left(g^{ij^{*}}D_{i}WD^{j*}W-3|W|^2\right),
\end{equation}
where $W$, the superpotential, is a holomorphic function of $a_{i}$ and
$D_{i}W\equiv \partial_{i}W+\partial_{i}K\,W$ and ${D_{a}}$ are
constants corresponding to a Fayet-Iliopoulos term. Assuming
$h_{ab}=\pm \delta_{ab}$, $v_{\mu}^{(a)}=0$ and $W=0$, action (\ref{1}) can be
obtained as a valid truncation of (\ref{2}). The validity can be checked
trivially by observing that the equations of motion of (\ref{2}) reduce to
those of (\ref{1}) with a cosmological constant term
$\frac{1}{2}D_{a}D^{a}$. Therefore all solutions to be
described in this paper can be seen as solutions of (\ref{2}).
\subsection{$N=1$, $d=4$ supergravity with gauged supermatter}
We have already remarked that $\bbbc P^{2}$ and $\bbbc P^{1}$ scalar
manifolds have isometry group of $SU(3)$ and $SU(2)$, and arise in the
$N=1$, $d=4$ supergravity with gauged supermatter action for these two
gauge groups. It is therefore natural to ask whether (\ref{1}) can be
obtained from $N=1$, $d=4$ supergravity with gauged supermatter action
. The bosonic part of the latter is\footnote{Here we have made a
choice for $h_{ab}$, which does not affect the ensuing arguments.}
\cite{WB}
\begin{equation}
{\mathcal I} =-\int d^4x\, \sqrt{g}\, \left(\frac{1}{2} R -
g_{ij^{*}}\tilde{\mathcal D}_{\mu}a^{i}\tilde{\mathcal D}^{\mu}a^{*j}
-\frac{1}{2} g^{2} {D^{(a)}}^{2} - \frac{1}{4}F^{(a)}_{\mu \nu}F^{\mu
\nu (a)} - {V}_{F} (a^{i},a^{*j})\right).\label{4}
\end{equation}
Here $\tilde{\mathcal D}_{\mu}a^{i}$ denotes
$\left({\partial}_{\mu}a^{(i)}-g v_{\mu}^{(a)} X^{i(a)} \right)$,
where $v_{\mu}^{(a)}$ is the multiplet of Yang-Mills potentials
($a=1,2,3$ for $SU(2)$ and $a=1,...,8$ for $SU(3)$) and $F^{(a)}_{\mu
\nu}$ are the Yang-Mills field strengths. The quantities $X^{i(a)}$,
together with their complex conjugates $X^{*j(a)}$, give the
holomorphic Killing vector fields
\begin{equation}
\begin{array}{rcl}
X^{(b)}&=&X^{i(b)}(a)\frac{\partial}{\partial a^{i}},\\
X^{*(b)}&=&X^{*i(b)}(a^{*})\frac{\partial}{\partial a^{*i}},
\end{array}
\end{equation}
corresponding to the infinitesimal isometries of the K\"{a}hler
manifold. The term ${D^{(a)}}^{2}$ gives a cosmological constant
$g^{2}/8$ for both gauge groups. The connection between the K\"{a}hler
scalar part of the theory and the gauge theory is that the isometry
group of the scalars is the gauge group of the full theory.

Obviously to obtain (\ref{1}) from (\ref{4}), we need the Yang-Mills
potentials as well as ${V}_{F} (a^{i},a^{*j})$ vanishing. The latter
is achieved by setting the superpotential to zero.  However, if one
sets $v_{\mu}^{(a)}=0$, it is not difficult to see that the equations
of motion for (\ref{4}) would not reduce to those of (\ref{1}).  This
is because the Yang-Mills fields are coupled to the scalar fields and
the corresponding equation of motion
\begin{equation}
{D}^{\mu}\!\left(\sqrt{g}F^{(a)}_{\mu\nu}\right)=g\,g_{ij*}\left(X^{(a)i} \tilde{\mathcal D}_{\mu}a^{*j}-\tilde{\mathcal D}_{\mu}a^{i}X^{*j(a)}\right)
\end{equation}
would not be satisfied for complex scalar fields in general. For this
to be possible one needs to set $g=0$, which takes us back to action
(\ref{2}).

In the following we adopt the conventions of \cite{MTW}, except that
we take $8 \pi G =1$. The Einstein equations are then
\begin{equation}
R_{\mu\nu}-\frac{1}{2} R g_{\mu\nu}+ \Lambda g_{\mu\nu}= T_{\mu\nu}.
\end{equation}
\section{$\bbbc P^2$: Field Equations and Solutions}\label{mean}
\subsection{Geometry of $\bbbc P^2$}
$\bbbc P^2$ is most simply described in terms of two complex scalar
coordinates, $a^{1}$ and $a^{2}$, such that the Hermitian metric
$g_{ij^{*}}$ is derived from the K\"{a}hler potential,
$K=\log{(1+a^{1}a^{*1}+a^{2}a^{*2})}$, as
\begin{equation}
g_{ij^{*}}= \partial^{2}K/\partial a^{i}\partial  a^{*j}
\end{equation}
As is well known (see for example, \cite{GP2}) $\bbbc P^2$ can also be
described in terms of four
real coordinates $(R, \Theta, \Psi, \Phi)$:
\begin{equation}
a^{1}=R \cos \frac{\Theta}{2} \exp\left(i \frac{\Psi + \Phi}{2}\right)
\end{equation}
and
\begin{equation}
a^{2}=R \sin\frac{\Theta}{2} \exp\left(i \frac{\Psi - \Phi}{2}\right),
\end{equation}
where
\begin{equation}
\begin{array}{rcl}
0&\leq&R\leq\infty\\
0&\leq&\Theta\leq\pi\\
0&\leq&\Phi\leq2\pi\\
0&\leq&\Psi\leq4\pi,\\
\end{array}
\end{equation}
giving the real Fubini-Study metric:
\begin{equation}
ds^2=
\frac{dR^2}{(1+\frac{\mu}{6}R^2)^{2}}+\frac{R^2}{4(1+\frac{\mu}{6}R^2)}(\sigma_{1}^{2}+\sigma_{2}^{2})
+\frac{R^2}{4(1+\frac{\mu}{6}R^2)^2}(\sigma_{3}^{2})\label{10}
\end{equation}
where $\sigma_{i}$ are the left-invariant one forms on $SU(2)$
(equivalently, on $S^{3}$):\\
\begin{equation}
\begin{array}{rcl}
\sigma_{1}&=\,&\cos{\Psi} d \Theta + \sin{\Theta}\sin{\Psi} d \Phi,\\
\sigma_{2}&=-&\sin{\Psi} d \Theta + \sin{\Theta}\cos{\Psi} d \Phi,\\
\sigma_{3}&=\,&\cos{\Theta}d \Phi + d \Psi\
\end{array}
\end{equation}
and obey the exterior algebra $d\sigma_{i}=-\frac{1}{2}\epsilon_{ijk}
\sigma_{j} \wedge \sigma_{k}$. The quantity $\mu$ is a positive
constant and is {\emph{not}} determined by the model.

One can check that metric (\ref{10}) is Einstein, i.e., that it satisfies the
Einstein equations with a (positive) cosmological constant ($R_{\mu \nu}=
\mu\, g_{\mu \nu}$) and hence $ \bbbc P^{2}$ is a compact
manifold. This can also be verified explicitly by
making the coordinate transformation $R= \sqrt{\frac{6}{\mu}}\tan\chi$ ($0\leq\chi\leq\frac{1}{2}\pi$):
\begin{equation}
ds^2=\frac{6}{\mu} \left(d\chi^{2}+
\frac{1}{4}\sin^{2}\chi(\sigma_{1}^{2}+\sigma_{2}^{2})
+\sin^{2}\chi\cos^{2}\chi (\sigma_{3}^{2}) \right).\\
\end{equation}
The constant-$R$ surfaces of (\ref{10}) are non-trivial $S^1$ bundles
over $S^2$ being invariant under the group action of $SU(2)\times
U(1)$. (This, however, is not the maximal symmetry of $ \bbbc P^{2}$.)
Since the periodicity of $\Psi$-coordinate is $4\pi$, they are
topologically $S^3$.  Near $R=0$ the metric approaches flat space and
near $R=\infty$ it collapses to a 2-sphere of finite radius. These
refer to the ``nut'' and ``bolt'' of the metric -- terms which will
be made clearer later in the paper. For the purpose of much of the
discussions below, we will conveniently set $\mu=6$ in (\ref{10})
unless this results in loss of generality. We will revert to the
general form (\ref{10}) where appropriate, as in the principal set of
explicit solutions obtained in this paper.
\subsection{Metric Ansatz for $\a^{4}$ and Field Equations}
For the metric on the Riemannian `spacetime' manifold $(\a^{4},
g_{\mu\nu})$, we take the following Ansatz relating the coordinates of
`spacetime' (here $r$, $\theta$, $\psi$, $\phi$) with those of the
target manifold:
\begin{equation}
\begin{array}{rcl}
R(x^{\mu})&=& R(r),\\
\\
\Theta(x^{\mu})=\theta, \Psi(x^{\mu})&=&\psi,\Phi(x^{\mu})=\phi.\label{13}
\end{array}
\end{equation}
One finds that:
\begin{equation}
\begin{array}{rcl}
2g_{ij^{*}}{\partial}_{r}a^{i}{\partial}_{r}a^{*j}&= &\frac{2
R'^{2}}{(1+R^{2})^{2}},\\
2g_{ij^{*}}{\partial}_{\theta}a^{i}{\partial}_{\theta}a^{*j}&=&\frac{1}{2}\frac{R^{2}}{(1+R^{2})},\\
2g_{ij^{*}}{\partial}_{\psi}a^{i}{\partial}_{\psi}a^{*j}&=&\frac{1}{2}\frac{R^{2}}{(1+R^{2})^{2}},\\
2g_{ij^{*}}{\partial}_{(\psi}a^{i}{\partial}_{\phi)}a^{*j}&=&\frac{1}{2}\frac{R^{2}}{(1+R^{2})^{2}}\cos\theta,\\
2g_{ij^{*}}{\partial}_{\phi}a^{i}{\partial}_{\phi}a^{*j}&=&\frac{1}{2}\frac{R^{2}}{(1+R^{2})^{2}}(1+R^2\sin^{2}\theta)\
\end{array}
\end{equation}
while all other components of
$2g_{ij^{*}}{\partial}_{(\mu}a^{i}{\partial}_{\nu)}a^{*j}$ are zero.
This naturally suggests the metric Ansatz for $\a^{4}$, of Riemannian
biaxial Bianchi-IX type:
\begin{equation}
ds^2= dr^2 + a^2(r)[(\sigma_{1})^{2}+(\sigma_{2})^2]+
b^2(r)(\sigma_{3})^{2}.
\end{equation}
Hence the non-zero components of $T_{\mu\nu}$ are:
\begin{equation}
\begin{array}{rcl}
T_{rr}&=&\frac{R'^{2}}{(1+R^{2})^{2}}-\left(\frac{R^{2}}{2(1+R^{2})
a^{2}}+\frac{R^{2}}{4(1+R^{2})^{2} b^{2}}\right),\\

\\
T_{\theta
\theta}&=&-\left(\frac{R'^{2}}{(1+R^{2})^{2}}+
\frac{R^{2}}{4(1+R^{2})^{2}b^{2}}\right) a^{2},\\

\\
T_{\psi
\psi}&=&-\left(\frac{R'^{2}}{(1+R^{2})^{2}}+\frac{R^{2}}{2(1+R^{2})a^{2}}-\frac{R^{2}}{4(1+R^{2})^{2}
b^{2}}\right) b^{2},\\

\\
T_{\psi \phi}&=&T_{\phi \psi}= T_{\psi \psi}\cos\theta,\\
\\
T_{\phi \phi}&=&T_{\theta \theta}\sin^{2}\theta+T_{\psi \psi}\cos^{2}\theta.\

\end{array}
\end{equation}
\subsection*{\it Scalar field equations}
Non-linear sigma models are special cases of harmonic maps from the
spacetime to the target manifold (see, for example, \cite{EL}). Harmonic maps are governed by the
equations:
\begin{equation}
\nabla^{\mu}\nabla_{\mu} X^{A}+ \Gamma^{A}_{BC}\nabla^{\mu}X^{B}\nabla_{\mu} X^{C}=0,\label{17}
\end{equation}
where the $X^{A}$ and $\Gamma^{A}_{BC}$ are respectively the
coordinates and Christoffel symbols of the target manifold. For the
present case, it is convenient to make use of the fact that $\bbbc
P^{2}$ can be given four real coordinates ($R, \Theta, \Psi, \Phi$)
and a real metric (\ref{10}). Instead of finding (complex) field
equations for $a^{i}$ and $a^{*j}$ by varying the action (\ref{1}),
and then translating them into real coordinates, we simply use
(\ref{17}) to find the field equations for the real coordinates ($R,
\Theta, \Psi, \Phi$) of the Fubini-Study metric. By the Ansatz (\ref{13}),
angular coordinates are determined by the corresponding `spacetime'
coordinates. The only non-trivial equation is for $R$: on using the
identity $\nabla^{\mu}\nabla_{\mu}=\frac{1}{\sqrt{g}}
\partial_{\mu}\left(\sqrt{g}\partial^{\mu}\right)$, we find the
equation for $R (r)$ (Eq.(\ref{19}) below).

The whole classical problem now reduces to that of finding solutions to the
following set of equations:
\begin{equation}
\begin{array}{rcl}
(\frac{a'}{a})^{2} + 2 \frac{a'}{a} \frac{b'}{b} + 
\frac{1}{4}\frac{b^{2}}{a^{4}}-\frac{1}{a^{2}}&=&\frac{R'^{2}}{(1+R^{2})^{2}}
-\left(\frac{R^{2}}{2
a^{2}(1+R^{2})}+\frac{R^{2}}{4 b^{2}(1+R^{2})^{2}}\right)-\Lambda,\\
\\
\frac{a''}{a} -\frac{a'}{a} \frac{b'}{b}
+\frac{1}{4}\frac{b^{2}}{a^{4}}&=&-\frac{R'^{2}}{(1+R^{2})^{2}}+\frac{R^{2}}{4
b^{2}(1+R^{2})^{2}},\\
\\
\frac{b''}{b} + 2 \frac{a'}{a}
\frac{b'}{b}-\frac{1}{2}\frac{b^{2}}{a^{4}}&=&- \frac{R^{2}}{2
b^{2}(1+R^{2})^{2}}-\Lambda,\label{18}
\end{array}
\end{equation}
with
\begin{equation}
R'' +\left(2\frac{a'}{a}+\frac{b'}{b}\right) R' -
\frac{2RR'^{2}}{(1+R^{2})}
-\frac{R}{2 a^{2}}
+ \frac{R(R^{2}-1)}{4(1+R^{2})b^{2}}=0.\\\label{19}
\end{equation}
The first-order constraint equation is consistent with the three other
second-order equations, as one can check. A typical solution $(a(r),
b(r), R(r))$ would involve numerical integration. One can check that
the full system of equations admits consistent (regular) power-series
solutions for $(a(r), b(r), R(r))$ near $r=0$. This can be used as the
starting-point in finding numerical solutions -- to be studied
elsewhere. In this paper we study some analytic solutions.

\section{Solutions}

\subsection{Einstein Metrics}
One may try to identify $R$ with $r$. This is similar to the approach
in \cite {GJOP} where the scalar coupling was initially taken to be
arbitrary, and $\bbbc P^{2}$ was found to be a solution for $\a^{4}$
for a particular value of the scalar coupling, leading to the concept
of spontaneous scalar compactification \cite{GZ, OP} -- a concept
which has since been used to compactify higher dimensional spacetimes
in various higher-dimensional theories.

However, in our case identifying $R$ with $r$ means that we have to
solve:
\begin{equation}
{R_{\mu\nu}}_{{\a}^4}=2 {g_{{\mu\nu}_{\bbbc P^{2}}}} + \Lambda {g_{\mu\nu}}_{{\a}^4}.\label{20}
\end{equation}
For $\Lambda=0$, it is not difficult to see that Eq.(\ref{20}) cannot
be satisfied by any Einstein metric, as follows. For an Einstein space
$R_{\mu\nu}=\lambda g_{\mu\nu}$ ($\lambda$ is a constant), which in
our case would mean ${g_{\mu\nu}}_{{\a}^4}= \frac {2}{\lambda}
g_{{\mu\nu}_{\bbbc P^{2}}}$, implying that the two metrics are related
by a constant conformal factor. This would imply that the Ricci
tensors are equal: $R_{{\mu\nu}_{{\a}^4}}=R_{{\mu\nu}_{\bbbc
P^{2}}}$. This gives $R_{{\mu\nu}_{\bbbc P^{2}}}=2 g_{{\mu\nu}_{\bbbc
P^{2}}}$ -- a contradiction, since in this case $R_{{\mu\nu}_{\bbbc
P^{2}}}=6 g_{{\mu\nu}_{\bbbc P^{2}}}$. However, this does not exhaust
the possibilities as we will see below.

First note that for the special value of the cosmological constant
$\Lambda=4$, we can recover the $\bbbc P^{2}$ as the solution for
${\a}^4$.  This generalizes directly. As remarked earlier, the $\mu$
for $\bbbc P^{2}$ is not determined by the theory. Identifying $R$
with $r$, the generalization of (\ref{20}) is:
\begin{equation}
{R_{\mu\nu}}_{{\a}^4}=\frac{12}{\mu} {g_{{\mu\nu}_{\bbbc P^{2}}}} + \Lambda {g_{\mu\nu}}_{{\a}^4}.
\end{equation}
This can be solved for $\mu$, and hence the corresponding Fubini-Study metric
satisfying (21) can be found. In other words, it is the reverse process: we
give the correct metric on $\bbbc P^{2}$ to get the same metric on
$\a^{4}$, i.e., we solve
\begin{equation}
\frac{12}{\mu}+\Lambda=\mu
\end{equation}
for $\mu$. This gives $\mu=  \frac{1}{2} (\Lambda +
\sqrt{\Lambda^{2}+48})$, for both positive and negative $\Lambda$;
hence the corresponding Fubini-Study metric on $\a^{4}$ can be
found. For $\Lambda=0$ this gives $\mu=2 \sqrt{3}$. 
\subsection{R= $constant$ solutions ?}
Clearly, in the case $R(r)=0$, the field equations reduce to those of a
biaxial Bianchi-IX model (without matter) admitting a $SU(2) \times
U(1)$ isometry group. The solutions are the general two-parameter
Riemannian Taub-NUT-(anti-)de Sitter family of metrics \cite{GP1}: 
\begin{equation}
ds^{2}=\frac{\rho^{2} - L^{2}}{\Delta} d\rho^{2}+ \frac{4
L^{2}\Delta}{\rho^{2}-L^{2}}(d\psi+\cos \theta
d\phi)^{2}+(\rho^{2}-L^{2})(d\theta^{2}+\sin^{2}\theta d\phi^{2}),\label{23}
\end{equation}
where
\begin{equation}
\Delta=\rho^{2}-2M\rho +L^{2}+\Lambda( L^{4}+2
L^{2}\rho^{2}-\frac{1}{3}\rho^{4}).
\end{equation}
This general form, however, is only valid for a coordinate patch for
which $\Delta \ne 0$.  At the roots, the metric degenerates to the
two-dimensional fixed-point set of the Killing vector field $\partial/
\partial \psi$; they are round 2-spheres of constant radii and have
been dubbed ``bolts'' \cite{GH}. However, if a root occurs at
$\rho=|L|$, the corresponding set of fixed points is zero-dimensional,
as the 2-sphere then collapses to a point; such a point is called a
``nut'' \cite{GH}. Such nuts and bolts are not necessarily regular
points of the metric. For them to be regular, the metric has to close
smoothly. This will be discussed in Section 5.1 in greater detail. In
general, one arrives at two one-parameter family of metrics, the
self-dual Taub-Nut-(anti-)de Sitter and the Taub-Bolt-(anti-)de Sitter
metrics, by imposing the condition of regularity. Known examples of
Bianchi-IX metrics and instantons arise as special cases of them. For
positive cosmological constant, known examples include the usual round
metric on $S^{4}$, the Fubini-Study metric on $\bbbc P^{2}$
\cite{GP2}. For vanishing cosmological constant, the known solutions
are the self-dual Taub-Nut instanton \cite{Hawk}, the Taub-Bolt
instanton \cite{Page}.  For negative cosmological constant, one
special solution is the Bergman metric on $\overline {\bbbc P^{2}}$,
which is just the Fubini-Study metric with the cosmological constant
reversed in sign\footnote{There are solutions, the Eguchi-Hanson
metrics \cite{EH} for example, whose level-surfaces are not
topologically $S^3$. Since we have taken the Ansatz in which the
$\psi$-coordinate of $\a^4$ has a period of $4\pi$, as the
$\Psi$-coordinate of the Fubini-Study metric (\ref{10}) of the target
manifold ${\bbbc P^{2}}$, such metrics are automatically
precluded. For more discussions on such metrics, see \cite{AkG}.}.

Although no new metric solutions have been obtained in this rather
trivial limit $R \equiv 0$, one can note that the scalar manifold is
crucial in fixing the symmetry of the hypersurface of $\a^{4}$ at
constant $r$ to be at least $SU(2) \times U(1)$-invariant, unlike the
case in which one ignores the internal geometry and just takes a more
\emph{ad hoc} Ansatz for $\a^{4}$. Thus, biaxial Bianchi-IX metrics
arise naturally by virtue of the ``hedgehog''-type Ansatz for the
scalar fields living on the internal space.  \\ One might next ask
whether any solutions exist for which $R(r)$ is a non-zero
constant. In this case the scalar field equation reads:
\begin{equation}
\frac{R}{2 a^{2}}- \frac{R(R^{2}-1)}{4(1+R^{2})b^{2}}=0
\end{equation}
so that $a$ and $b$ are proportional:
\begin{equation}
a=\sqrt {2}\sqrt { {\frac {{R}^{2}+1}{{R}^{2}-1}}}\,b \label{26}
\end{equation}
with $R^{2} > 1$. To examine the existence of such a solution, write the two evolution
equations (\ref{18}) in a slightly
different way:
\begin{equation}
\begin{array}{rcl}
\frac{a''}{a} -\frac{a'}{a} \frac{b'}{b}
&=&-\frac{1}{4}\frac{b^{2}}{a^{4}}+\frac{R^{2}}{4
b^{2}(1+R^{2})^{2}},\\
\\
\frac{b''}{b} -(\frac{a'}{a})^{2}&=&\frac{3}{4}\frac{b^{2}}{a^{4}}-\frac{1}{a^{2}}+\frac{R^{2}}{2
a^{2}(1+R^{2})}- \frac{R^{2}}{4
b^{2}(1+R^{2})^{2}}.\\
\end{array}
\end{equation}
For $a$ and $b$ proportional to each other, both right hand sides
should be identical.  But, on substituting (\ref{26}), one finds that this
requires $R=3/5$, contradicting the requirement $R^{2} > 1$. One might
then have thought that there are no other geometrically significant
solutions corresponding to a fixed value of $R$ and solving the
coupled system of equations (\ref{18}) and (\ref{19}), except for $R
\equiv 0$. However, this is not so, as we have so far omitted the case
`$R \equiv \infty$'.
\subsubsection{R$=\infty$} 
One may investigate the neighbourhood of $R
\rightarrow \infty$ by defining $u(r)=\frac{1}{R(r)}$. The field
equations become:
\begin{equation}
\begin{array}{rcl}
(\frac{a'}{a})^{2} + 2 \frac{a'}{a} \frac{b'}{b} +
\frac{1}{4}\frac{b^{2}}{a^{4}}-\frac{1}{a^{2}}&=&\frac{u'^{2}}{(1+u^{2})^{2}}-
\left(\frac{1}{2(1+u^{2})
a^{2}}+\frac{u^{2}}{4 b^{2}(1+u^{2})^{2}}\right)-\Lambda,\\
\\
\frac{a''}{a} -\frac{a'}{a} \frac{b'}{b}
+\frac{1}{4}\frac{b^{2}}{a^{4}}&=&-\frac{u'^{2}}{(1+u^{2})^{2}}+\frac{u^{2}}{4
b^{2}(1+u^{2})^{2}},\\
\\
\frac{b''}{b} + 2 \frac{a'}{a}
\frac{b'}{b}-\frac{1}{2}\frac{b^{2}}{a^{4}}&=&- \frac{u^{2}}{2
b^{2}(1+u^{2})^{2}}-\Lambda,
\end{array}
\end{equation}
with
\begin{equation}
u'' +\left(2\frac{a'}{a}+\frac{b'}{b}\right) u' -
\frac{2uu'^{2}}{(1+u^{2})}
+\frac{u}{2 a^{2}}\\
- \frac{u(u^{2}-1)}{4(1+u^{2})b^{2}}=0.\\
\end{equation}
It is clearly consistent to set $u=0$ (corresponding
to $R=\infty$). However, in contrast to the
$R(r)=0$ case, here ($R=\infty$) the energy-momentum tensor is
non-zero. This is due to the fact that at $R=0$, $ \bbbc
P^{2}$ degenerates to a point (a ``nut'') whereas at $R=\infty$ it
degenerates to an $S^{2}$ of constant radius (a ``bolt'' -- as in
section $4$). The
three field equations read:
\begin{equation}
\left(\frac{a'}{a}\right)^{2} + 2 \frac{a'}{a} \frac{b'}{b} +
\frac{1}{4}\frac{b^{2}}{a^{4}}-\frac{1}{a^{2}}=-\frac{1}{2
a^{2}}-\Lambda,\label{30}
\end{equation}
\begin{equation}
\frac{a''}{a} -\frac{a'}{a} \frac{b'}{b}
+\frac{1}{4}\frac{b^{2}}{a^{4}}=0,
\end{equation}
\begin{equation}
\frac{b''}{b} + 2 \frac{a'}{a}
\frac{b'}{b}-\frac{1}{2}\frac{b^{2}}{a^{4}}= -\Lambda.
\end{equation}
These equations are just as in the biaxial Bianchi-IX case with only a
cosmological constant (and no matter), except for the presence of the
$\frac{1}{2 a^{2}}$ term in (\ref{30}). However, by the rescaling
$\alpha(r)=\sqrt{2}\,a(r)$ and $\beta(r)= 2\, b(r)$, these equations
reduce to those for the Bianchi-IX case with just a cosmological
constant $\Lambda$:
\begin{equation}
\left(\frac{a'}{a}\right)^{2} + 2 \frac{a'}{a} \frac{b'}{b} +
\frac{1}{4}\frac{b^{2}}{a^{4}}-\frac{1}{a^{2}}=-\frac{1}{2
a^{2}}-\Lambda,\label{6.33}
\end{equation}
\begin{equation}
\frac{a''}{a} -\frac{a'}{a} \frac{b'}{b}
+\frac{1}{4}\frac{b^{2}}{a^{4}}=0,
\end{equation}
\begin{equation}
\frac{b''}{b} + 2 \frac{a'}{a}
\frac{b'}{b}-\frac{1}{2}\frac{b^{2}}{a^{4}}= -\Lambda.
\end{equation}
Therefore \emph{all} our solutions with $R= \infty$ can be put into
1-1 correspondence with those of the Einstein-$\Lambda$ system without
matter. The two scale factors $a(r)$ and $b(r)$ are dilated by factors
of $1/\sqrt{2}$ and $\frac{1}{2}$ when compared with the biaxial
Bianchi IX solution with only a $\Lambda$ term.

The general solutions of Bianchi-IX type, obeying the Einstein
equations with a $\Lambda$ term, are the Taub-NUT family of metrics
(\ref{23}) as discussed already.  Therefore the solutions to the case
`$R=\infty$' are given by the ``extended'' metrics:
\begin{equation}
ds^{2}=\frac{\rho^{2} - L^{2}}{\Delta} d\rho^{2}+ \frac{4
L^{2}\Delta}{\rho^{2}-L^{2}}\left(1-\frac{3}{\mu}\right)^{2}(d\psi+\cos \theta
d\phi)^{2}+(\rho^{2}-L^{2})\left(1-\frac{3}{\mu}\right)(d\theta^{2}+\sin^{2}\theta
d\phi^{2})\label{36}
\end{equation}
This is a two-parameter family of metrics which clearly are in 1-1
correspondence with their no-matter counterparts. We now discuss the
regularity of these metrics, and then show
how the same set of solutions is allowed for the $\bbbc P^{1}$
sigma-model coupled to gravity.
\section{New Metrics and their Regularity}
\noindent
It is convenient to rewrite the two-parameter family of metrics (\ref{36})
in the form:
\\
\begin{equation}
ds^{2}=\frac{\zeta^{2} - l^{2}}{\tilde{\Delta} (1-\frac{3}{\mu})}
d\zeta^{2}+ \frac{4
l^{2}\tilde{\Delta}(1-\frac{3}{\mu})}{\zeta^{2}-l^{2}}(d\psi+\cos
\theta d\phi)^{2}+(\zeta^{2}-l^{2})(d\theta^{2}+\sin^{2}\theta
d\phi^{2}),
\end{equation}
where
\begin{equation}
\tilde{\Delta}=\zeta^{2}-2m\rho
+l^{2}+\frac{\Lambda}{(1-\frac{3}{\mu})}( l^{4}+2
l^{2}\zeta^{2}-\frac{1}{3}\zeta^{4}).
\end{equation}
The lower-case quantities $m$, $l$ are continuous parameters, related
to those of (\ref{36}) by $m=\sqrt{(1-\frac{3}{\mu})}\,M$ and
$l=\sqrt{(1-\frac{3}{\mu})}\,L$. Again, this metric is only valid for
a coordinate patch for which $\tilde{\Delta} \ne 0$ ($\tilde{\Delta}$
having four roots).
\subsection{Regularity of the Taub-NUT-(anti-)de Sitter family}
As already remarked in Section 4.2, the four roots of $\Delta=0$ are
not in general regular points of the metric (\ref{23}). In this
section we briefly describe how one obtains two one-parameter family
of metrics from (\ref{23}) by making one of the roots regular (for
more details, see \cite{Akbar,mma,AkG,Chamblin}). The condition of
regularity of (\ref{23}) at any point $\rho_{bolt}$ where $\Delta=0$
works out to be \cite{Page}:
\begin{equation}
\frac{d}{d\rho}\left(\frac{\Delta}{\rho^2-L^2}
\right)_{(\rho=\rho_{root})}=\frac{1}{2L}
\end{equation}
which amounts to imposing a relation between $M$ and $L$. Thus the
condition of regularity reduces the two-parameter Taub-NUT-(anti-)de
Sitter family essentially to one-parameter families.
\subsubsection*{\it Self-dual Taub-Nut-(anti-)de Sitter}
The metric (\ref{23}) has a nut if $\Delta=0$ at $\rho=|L|$. This
means:
\begin{equation}
M=  L\left(1+\frac{4}{3}\Lambda^{2}\right)\label{40}
\end{equation}
which is also the condition of self-duality of the Weyl tensor of the
metric (\ref{23}) \cite{mma,GP2}. Thus, the condition of regularity
provides precisely the relation between the two parameters ($L$ and
$M$) such that the metric has a (anti-)self-dual Weyl tensor. Assuming
this relation (\ref{40}), one finds:
\begin{equation}
\Delta=(\rho - L)^{2}-\frac{1}{3}\Lambda (\rho+3L)(\rho-L)^{3}.
\end{equation}
It is easy to see that the condition of regularity is automatically
fulfilled.
\subsubsection*{\it Taub-Bolt-(anti-)de Sitter}
If $\Delta=0$ has a root at $\rho\ne|L|$, then the set of fixed points
of $\partial/\partial \psi$ is necessarily a two dimensional bolt. If
the bolt is at $\rho_{bolt}$, one has:
\begin{equation}
M=\frac{1}{6}\,{\frac {3\,{\rho_{bolt}}^{2}-\Lambda\,{\rho_{bolt}}^{4}+3\,{L}^{2}+3\,\Lambda
\,{L}^{4}+6\,\Lambda\,{L}^{2}{\rho_{bolt}}^{2}}{\rho_{bolt}}}
\end{equation}
The condition of regularity then reads:
\begin{equation}
{\frac
{-\Lambda\,{\rho_{bolt}}^{2}+{L}^{2}\Lambda+1}{\rho_{bolt}}}=\frac{1}{2L},\label{43}
\end{equation}
which requires $L< \rho_{bolt}<2L$ in the case of positive cosmological
constant and $\rho_{bolt} > 2L$ for negative cosmological constant, since
$\rho > L$ for $L$ positive. Eq.(\ref{43}) can be solved to locate the bolt which is at
\begin{equation}
\rho_{bolt}=\frac{1}{4}\,\frac {-1+\sqrt
{1+16\,\Lambda^{2}L^{4}+16\,L^{2} \Lambda\,}}{\Lambda L}
\end{equation}
for positive cosmological constant, whence
\begin{equation}
M={\frac {1}{96}}\,{\frac {1+\sqrt {1+16\,{\Lambda}^{2}{L}^{4}+16\,\Lambda
\,{L}^{2}}\left (8\,\Lambda\,{L}^{2}+32\,{\Lambda}^{2}{L}^{4}-1\right 
)}{{\Lambda}^{2}{L}^{3}}}.
\end{equation}
However, when the cosmological constant is negative (written as $\Lambda \equiv
-\lambda$), the bolt would be either at
\begin{equation}
\rho_{bolt}=\frac{1}{4}\,{\frac {1-\sqrt {1+16\,{\lambda}^{2}{L}^{4}-16\,{L}^{2}
\lambda}}{\lambda\,L}}
\end{equation}
or at
\begin{equation}
\rho_{bolt}=\frac{1}{4}\,{\frac {1+\sqrt {1+16\,{\lambda}^{2}{
L}^{4}-16\,{L}^{2}\lambda}}{\lambda\,L}}
\end{equation}
provided that the quantity under the square root is non-negative. This
last requirement, together with that of $\rho_{bolt}>2L$, restricts
$L$:
\begin{equation}
\lambda\,L^{2}\le \left(\frac{1}{2}-\frac{\sqrt{3}}{4}\right)\,\,\,\, (\sim 0.066987298)
\end{equation}
Therefore \emph{only} for this range of $L$ can one get a regular
bolt, and $M$ is:
\begin{equation}
M= {\frac {1}{96}}\,{\frac {1 \pm \sqrt {1+16\,{\lambda}^{2}{L}^{4}-16\,\lambda
\,{L}^{2}}\left (32\,{\lambda}^{2}{L}^{4}-8\,\lambda\,{L}^{2}-1\right 
)}{{\lambda}^{2}{L}^{3}}}.
\end{equation}
The positive and negative signs correspond to the first and second
values of
$\rho_{bolt}$ above, respectively. Hence, for an $L$ which is in the permissible range, there are two
choices which give a regular bolt, depending on the choice of $M$.\\
\subsection{The Nuts and Bolts of the new Solutions}
Having recalled the above properties of the Taub-Nut/Bolt-(anti-)de
Sitter family of metrics, we can now analyze our solutions
systematically. The condition for regularity for the metric at the
bolt in this case is:
\begin{equation}
\left(1-\frac{3}{\mu}\right)\frac{d}{d\zeta}\left(\frac{\tilde{\Delta}}{\zeta^2-l^2}
\right)_{(\zeta=\zeta_{root}=l)}=\frac{1}{2l}.\label{54}
\end{equation}
As before this works as a necessary condition for regularity near a
nut. However, near the nut one requires the metric to approach the
flat metric on ${\Bbb E}^4$ which is not guaranteed {\it{a priori}} by it.
\subsection*{No Regular Nuts}
It is fairly straightforward to see that there will be no regular nut solutions. For a nut, $\tilde{\Delta}$ will have a root at $\zeta=l$, which
implies that
\begin{equation}
\tilde{\Delta}=(\zeta - l)^{2}-\frac{1}{3}\frac{\Lambda}{(1-\frac{3}{\mu})} (\zeta+3 l)(\zeta-l)^{3}.
\end{equation}
It is easy to check that the only way to satisfy condition (\ref{54})
is by taking $\mu$ to infinity. Note that, as $\zeta \rightarrow l$,
the terms involving $\Lambda$ in $\tilde{\Delta}$ fall faster than the
other terms not involving $\Lambda$. Hence, near $\zeta = l$, the
metric is the matter-equivalent to the self-dual Taub-Nut instanton
\cite {Hawk}:
\begin{equation}
ds^2=\frac{1}{(1-\frac{3}{\mu})}\left(\frac{\zeta + l}{\zeta - l}\right) d\zeta^{2}+4(1-\frac{3}{\mu}) l^{2}\left(\frac{\zeta - l}{\zeta + l}\right) (d\psi+\cos \theta
d\phi)^{2}+(\zeta^{2}-l^{2})(d\theta^{2}+\sin^{2}\theta d\phi^{2})\label{55}
\end{equation}
When the cosmological constant is zero, this is the metric away from
the nut as well. In any case it is straightforward to check directly
that the metric (\ref{55}) can not approach flat metric near the nut. Thus,
it is not possible to have a regular nut solution of this
type. However, the situation with bolts is different.\\
\subsection*{The Regular Bolts}
For the following computations, we set $s=(1-\frac{3}{\mu})$. The metric then becomes:
\begin{equation}
ds^{2}=\frac{\zeta^{2} - l^{2}}{\tilde{\Delta}\, s } d\zeta^{2}+ \frac{4
l^{2}\tilde{\Delta}\,s }{\zeta^{2}-l^{2}}(d\psi+\cos \theta
d\phi)^{2}+(\zeta^{2}-l^{2})(d\theta^{2}+\sin^{2}\theta
d\phi^{2}),
\end{equation}
where
\begin{equation}
\tilde{\Delta}=\zeta^{2}-2m\,\zeta +l^{2}+\frac{\Lambda}{s}\,(\, l^{4}+2
l^{2}\zeta^{2}-\frac{1}{3}\zeta^{4}).\label{5.18}
\end{equation}
The regularity condition reads:
\begin{equation}
s\,\, \frac{d}{d\zeta}\left(\frac{\tilde{\Delta}}{\zeta^2-l^2}
\right)_{(\zeta=\zeta_{bolt})}=\frac{1}{2l}\,\,,
\end{equation}
which for positive cosmological constant reads:
\begin{equation}
2\,\Lambda\,l{\zeta_{bolt}}^{2}+\zeta_{bolt}-2\,\Lambda\,{l}^{3}-2\,s\,l=0,
\end{equation}
and hence requires $l<\zeta_{bolt}<2\,s\,l$. This shows that one must
have $s > 1/2$ to have any regular bolt solution (as $\zeta \ge l$) in
the case of a positive cosmological constant. Note that, in the
limiting case $s=1/2$, it is \emph{not} possible to have a regular
bolt. However, when $s > 1/2$, the bolt is at:
\begin{equation}
\zeta_{bolt}=\frac{1}{4}\,\frac {-1+\sqrt
{1+16\,\Lambda^{2}l^{4}+16\,l^{2} \Lambda\,s}}{\Lambda l}.
\end{equation}
In the case of a negative cosmological constant, $\Lambda \equiv
-\lambda$, the situation becomes much more
interesting.
The regularity condition then reads:
\begin{equation}
{\frac {\lambda\,{\zeta_{bolt}}^{2}-{l}^{2}\lambda+s}{\zeta_{bolt}}}=\frac{1}{2l},
\end{equation}
which will require $\zeta_{bolt} > 2\,l\,s$ in this case, since $\zeta
> l$. This limits $s$ to be greater than or equal to $1/2$. The bolts
are located at
\begin{equation}
\zeta_{bolt}=\frac{1}{4}\,{\frac {1-\sqrt {1+16\,{\lambda}^{2}{l}^{4}-16\,\lambda{l}^{2}{s}}}{\lambda\,l}}\label{62}
\end{equation}
or
\begin{equation}
\zeta_{bolt}=\frac{1}{4}\,{\frac {1+\sqrt {1+16\,{\lambda}^{2}{l}^{4}-16\,\lambda{l}^{2}{s}}}{\lambda\,l}}.\label{63}
\end{equation}
These correspond to the two different ``choices'' of $m$ and hence would
not appear together in the same metric.

The quantity under the square root in Eqs.(\ref{62}) and (\ref{63}) should be
non-negative. Together with the restriction $\zeta_{bolt}>2\,l\,s$,
this puts a limit to the range of $l$:
\begin{equation}
\lambda\,l^{2}\le \frac{1}{2}\,s-\frac{1}{4}\sqrt{(4 s^{2}-1)}.
\end{equation}
For the limiting case $s=1/2$, there can hence only be one regular bolt
when the cosmological constant is negative. In this case the bolt is
located at:
\begin{equation}
\zeta_{bolt}=\frac{1}{4}\,\frac {2-4\,l^{2}\,\lambda}{\lambda\,l}
\end{equation}
which implies the restriction:
\begin{equation}
l^{2}\lambda < \frac{1}{4}
\end{equation}
(The other possibility is discarded as it places the bolt at
$\zeta=l$.) One can now calculate $m$ to be:
\begin{equation}
m=\frac{1}{24}\,{\frac
{64\,{l}^{6}{\lambda}^{3}-24\,{\lambda}^{2}{l}^{4}+1}{{l}
^{3}{\lambda}^{2}}}
\end{equation}
Therefore the one-parameter family of metrics satisfying the Einstein
equations with a negative cosmological constant and scalar field on a
``unit'' $\bbbc P^{2}$ with scalar self-coupling equal to $\frac{1}{2}$ is:
\begin{equation}
ds^{2}=\frac{\zeta^{2} - l^{2}}{F}\, d\zeta^{2}+ \frac{4
l^{2}\,F}{\zeta^{2}-l^{2}}\,(d\psi+\cos \theta
d\phi)^{2}+(\zeta^{2}-l^{2})\,(d\theta^{2}+\sin^{2}\theta
d\phi^{2}),
\end{equation}
where
\begin{equation}
F=\frac{1}{24}\,{\frac {\left (2\,\lambda\,l\zeta+2\,\lambda\,{l}^{2}-1\right 
)\left (4\,{l}^{2}{\lambda}^{2}{\zeta}^{3}-4\,{\zeta}^{2}{l}^{3}{
\lambda}^{2}+2\,l\lambda\,{\zeta}^{2}-20\,\zeta\,{\lambda}^{2}{l}^{4}+
2\,\lambda\,{l}^{2}\zeta+\zeta-12\,{l}^{5}{\lambda}^{2}\right )}{{l}^{
3}{\lambda}^{2}}}.
\end{equation}
$\zeta$ starts from $\frac{1}{4}\,\frac
{2-4\,l^{2}\,\lambda}{\lambda\,l}$ and goes to infinity; it is regular
everywhere provided that $l^{2}\lambda < \frac{1}{4}$. For values of
$s>\frac{1}{2}$, the two solutions are found by substituting the two
values of $m$ in (\ref{5.18}).
\subsection{$\bbbc P^{1}$ sigma-model}
The scalar manifold of $\bbbc P^{1}$ has one complex scalar
coordinate, $a$, and the metric
\begin{equation}
ds^{2}_{\bbbc P^{1}}=\frac{{da}{da^{*}}}{(1+aa^{*})^{2}}.\label{cpone}
\end{equation}
By the coordinate transformation
\begin{equation}
a= \tan \left(\frac{\Theta}{2}\right) \exp^{i \Phi},
\end{equation}
where $0 \leq \Theta \leq \pi$ and $0 \leq \Phi\leq2\pi$,  $\bbbc
P^{1}$ can
be given the real metric
\begin{equation}
ds^{2}_{\bbbc P^{1}}=\frac{1}{4}(d \Theta^{2}+\sin^{2} \Theta d \Phi^{2}).\label{72}
\end{equation}
This is the standard round metric on $S^{2}$ which satisfies the
Einstein equation with a cosmological constant of $1/4$.  Note that, as in the
case of the Fubini-Study metric (\ref{10}) on $\bbbc P^{2}$, the
metric (\ref{cpone}) is defined up to a positive arbitrary
constant. In the case of $\bbbc P^{2}$, the Fubini-Study
metric on $\bbbc P^{2}$ collapses to a ``bolt'' ($S^{2}$ of constant
radius) with the same metric as (\ref{72}) at $R=\infty$ (cf
(\ref{10})). Therefore, the $R=\infty$ solutions for $\bbbc P^{2}$ can
all be obtained from the $\bbbc P^{1}$ case by taking scalar Ansatz
$\Phi=\phi, \Theta=\theta$.
\section{Conclusion}
In this paper we have found some Euclidean solutions to the Einstein
field equations when $\bbbc P^{1}$ and $\bbbc P^{2}$-sigma-models are
coupled to gravity. Such manifolds arise as scalar manifolds in
supergravity plus supermatter Lagrangians. In the case of $\bbbc
P^{2}$, the sigma-model treatment is linked to the symmetry of
spacetime, here given by the group $SU(2) \times U(1)$. Among the
solutions obtained in this paper, there is a special class which are
in 1-1 correspondence with the two-parameter Taub-NUT-(anti-)de Sitter
family of metrics, and can be put in closed form. They exist as a
result of the non-trivial topology of $\bbbc P^{2}$ and can have
bolt-regularity -- however, the ``distortion'' made by the presence of
matter prohibits the possibility of any nut-type regularity. Finally,
we have shown how these metrics are also obtainable by coupling $\bbbc
P^{1}$ to gravity.

All solutions and much of the analysis in this paper goes through
equally to the Lorentzian r\'{e}gime if one takes an Ansatz of the
form $R=R(t)$. The analogous special-case solutions are readily
obtainable by taking $r \rightarrow it$ and are all in 1-1
correspondence with Lorentzian Taub-NUT-(anti-)de Sitter
spacetimes. The dynamics of the general field equations, i.e., the `$r
\rightarrow it$'-version of Eq.(\ref{18})-(\ref{19}), is left for future
investigation.
\section*{Acknowledgements}
We would like to thank Gary Gibbons, Carlos Nu\~{n}ez, Fernando
Quevedo, Susha Parameswaran and John Stewart for helpful discussions
and comments. MMA was supported by awards from the Cambridge
Commonwealth Trust, the Overseas Research Scheme and by DAMTP.

\end{document}